\documentclass[conference]{IEEEtran}
\usepackage{cite}

\usepackage[pdftex]{graphicx}
\usepackage{subfig}

\usepackage{smartdiagram}
\usesmartdiagramlibrary{additions}
\usepackage{forest}
\usepackage{amsmath}
\usepackage{hyperref}
\usepackage{url}

\usepackage{tikz}
\usetikzlibrary{shapes.geometric, arrows}

\usepackage{multirow}

\tikzstyle{raw} = [rectangle, rounded corners, minimum width=1.5cm, minimum height=1cm,text centered, draw=black, fill=black!30]
\tikzstyle{ensemble} = [rectangle, rounded corners, minimum width=2cm, minimum height=1cm,text centered, draw=black, fill=green!30]
\tikzstyle{smallcircle} = [circle, rounded corners, minimum width=0.05cm, minimum height=0.05cm,text centered, draw=black, fill=black!30]
\tikzstyle{file} = [trapezium, text centered, draw=black, trapezium right angle=130, fill=orange!30]

\tikzstyle{arrow} = [thick,->,>=stealth]

\begin{document}

\title{Efficient loading of reduced data ensembles produced at ORNL SNS/HFIR neutron time-of-flight facilities}

\author{
\IEEEauthorblockN{William F Godoy}
\IEEEauthorblockA{
\textit{Computer Science and}\\
\textit{Mathematics Division,}\\
\textit{Oak Ridge National Laboratory}\\
Oak Ridge, TN, USA\\
Email: godoywf@ornl.gov}
\and
\IEEEauthorblockN{Andrei T Savici}
\IEEEauthorblockA{
\textit{Neutron Scattering Division,}\\
\textit{Oak Ridge National Laboratory}\\
Oak Ridge, TN, USA\\
Email: saviciat@ornl.gov}
\and
\IEEEauthorblockN{Steven E Hahn}
\IEEEauthorblockA{
\textit{Computer Science and}\\
\textit{Mathematics Division,}\\
\textit{Oak Ridge National Laboratory}\\
Oak Ridge, TN, USA\\
Email: hahnse@ornl.gov}
\and
\IEEEauthorblockN{Peter F Peterson}
\IEEEauthorblockA{
\textit{Computer Science and}\\
\textit{Mathematics Division,}\\
\textit{Oak Ridge National Laboratory}\\
Oak Ridge, TN, USA\\
Email: petersonpf@ornl.gov}
}
\maketitle

\begin{abstract}
We present algorithmic improvements to the loading operations of certain reduced data ensembles produced from neutron scattering experiments at Oak Ridge National Laboratory (ORNL) facilities. 
Ensembles from multiple measurements are required to cover a wide range of the phase space of a sample material of interest.
They are stored using the standard NeXus schema on individual HDF5 files.
This makes it a scalability challenge, as the number of experiments stored increases in a single ensemble file.
The present work follows up on our previous efforts on data management algorithms, to address identified input output (I/O) bottlenecks in Mantid, an open-source data analysis framework used across several neutron science facilities around the world.
We reuse an in-memory binary-tree metadata index that resembles data access patterns, to provide a scalable search and extraction mechanism.
In addition, several memory operations are refactored and optimized for the current common use cases, ranging most frequently from 10 to 180, and up to 360 separate measurement configurations.
Results from this work show consistent speed ups in wall-clock time on the Mantid LoadMD routine, ranging from 19\% to 23\% on average, on ORNL production computing systems.
The latter depends on the complexity of the targeted instrument-specific data and the system I/O and compute variability for the shared computational resources available to users of ORNL's Spallation Neutron Source (SNS) and the High Flux Isotope Reactor (HFIR) instruments.
Nevertheless, we continue to highlight the need for more research to address reduction challenges as experimental data volumes, user time and processing costs increase.
\end{abstract}


\IEEEpeerreviewmaketitle

\section{Introduction}
Oak Ridge National Laboratory (ORNL) neutron science facilities, the Spallation Neutron Source (SNS) and the High-Flux Isotope Reactor (HFIR), produce large amounts of experimental raw data from a variety of instruments~\cite{Neutrons}. Raw data at SNS and HFIR is stored in ``event mode"~\cite{Granroth:ut5002}, in which each neutron is tagged with its arrival time, thus recording what is known as "time-of-flight" information.
In addition, detector and measurement characteristics annotations must be stored, to capture meaningful physical information under certain conditions~\cite{PETERSON201524}.
The result is a large database stored using the metadata-rich standard NeXus schema~\cite{Konnecke:po5029}, built on top of the self-describing hierarchical data format, HDF5~\cite{hdf5}, hosted at ORNL's SNS and HFIR computing facilities~\cite{Campbell_2010}. 
Raw event data needs to be post-processed into higher-level data products, to extract physical quantities of interest. Moreover, single experimental runs only expose limited information on complex material physics studied with neutron scattering techniques. Hence, experiments from a particular instrument need to cover a wide range of observational parameters ({\it e.g.} angular coverage, different temperatures or magnetic fields, etc.), which could also incur in significant overlap~\cite{doi:10.1063/1.4870050}. The latter results in large data set ensembles of different statistical significance, posing major challenges in post-processing and interpretation workflows.

In this paper, we tackle the data loading challenges for single-crystal inelastic experiments, which require many sample orientations~\cite{Horace}. The raw data and metadata from each orientation is pre-processed into a single ``multidimensional" data set ensemble. It contains all the information required to visualize different slices of the dynamic structure factor, typical in neutron scattering techniques. Loading this single, common, ``multidimensional" data set from disk and processing the results into a reduced data structure in memory for slicing the entire experimental domain is a major bottleneck in the analysis workflows at SNS and HFIR available to users.  

The Mantid framework~\cite{ARNOLD2014156} is an international collaboration between several neutron sciences facilities around the world; including ORNL’s SNS and HFIR~\cite{Neutrons}, the ISIS Neutron and Muons Source~\cite{THOMASON201961}, and The Institut Laue–Langevin (ILL)~\cite{AGERON1989197}. Mantid is used by several production workflows for data reduction and analysis at these facilities.  The present work is the result of the optimizations applied to the Mantid loading algorithm for multidimensional data ensembles named ``LoadMD"~\cite{ARNOLD2014156}. The latter has been identified as a bottleneck in the process of data visualization for single crystal experiments. Two major changes are applied to this algorithm, the first one is the caching of an ``in-memory index"~\cite{8990452} data structure, which reuses similar strategies outlined in our previous work addressing raw neutron data reduction bottlenecks\cite{9377836}; the second change is the reduction of ``LoadMD" memory allocation requests which, in a shared computational resource, can lead to increased variability in the overall loading operation. Impact is obtained from the wall-clock time scalability of ``LoadMD" operations on SNS/HFIR production computational systems\footnote{\label{note1}\url{https://analysis.sns.gov}} as a function of the number of orientations, between 10 to 360 experimental measurements, stored in data sets ensembles that are typical of the targeted applications~\cite{doi:10.1063/1.4870050}.

The paper is organized as follows, Section~\ref{sec:neutron_data_ensembles} provides a description of the neutron multidimensional data ensembles produced from raw event-based data and the loading and processing workflow for analyzing a full experiment. Section~\ref{sec:proposed_methodology} describes the proposed methodology to address the identified current bottleneck in loading operations from profiling the current operations in ``LoadMD", while Section~\ref{sec:results} presents results on SNS/HFIR data analysis computing systems, including the expected speed ups, scalability from comparing actual experimental data set sizes, and the variability in wall-clock times due to the shared-resource nature of the system. Finally, Section~\ref{sec:conclusions} presents the conclusions from our study outlining the need to continue improving existing data reduction workflows as data volumes continue to increase in future generation instruments at ORNL, thus resulting in a wealth of research opportunities for tackling scientific data challenges~\cite{8416399}.

\section{Neutron Multidimensional Data Ensembles} \label{sec:neutron_data_ensembles}
\begin{figure}[!h]
\centering
    
\begin{tikzpicture}[node distance=1.2cm]
\node (exp1)[raw]{EventWorkspace1};
\node (exp2)[raw, below of=exp1]{EventWorkspace2};
\node (exp3)[smallcircle, below of=exp2, yshift=-0.1cm]{...};
\node (expN)[raw, below of=exp3]{EventWorkspaceN};

\node (MD)[ensemble, right of=exp1, xshift=2cm]{MDWorkspace};
\node (MDFile)[file, below of=MD, xshift=1.5cm]{MDFile.nxs.h5};
\node (MDNew)[ensemble, below of=MDFile, yshift=-0.5cm]{MDWorkspace};

\node (slice)[below of=MDNew,yshift=-0.55cm]{\includegraphics[width=.1\textwidth,height=.07\textwidth]{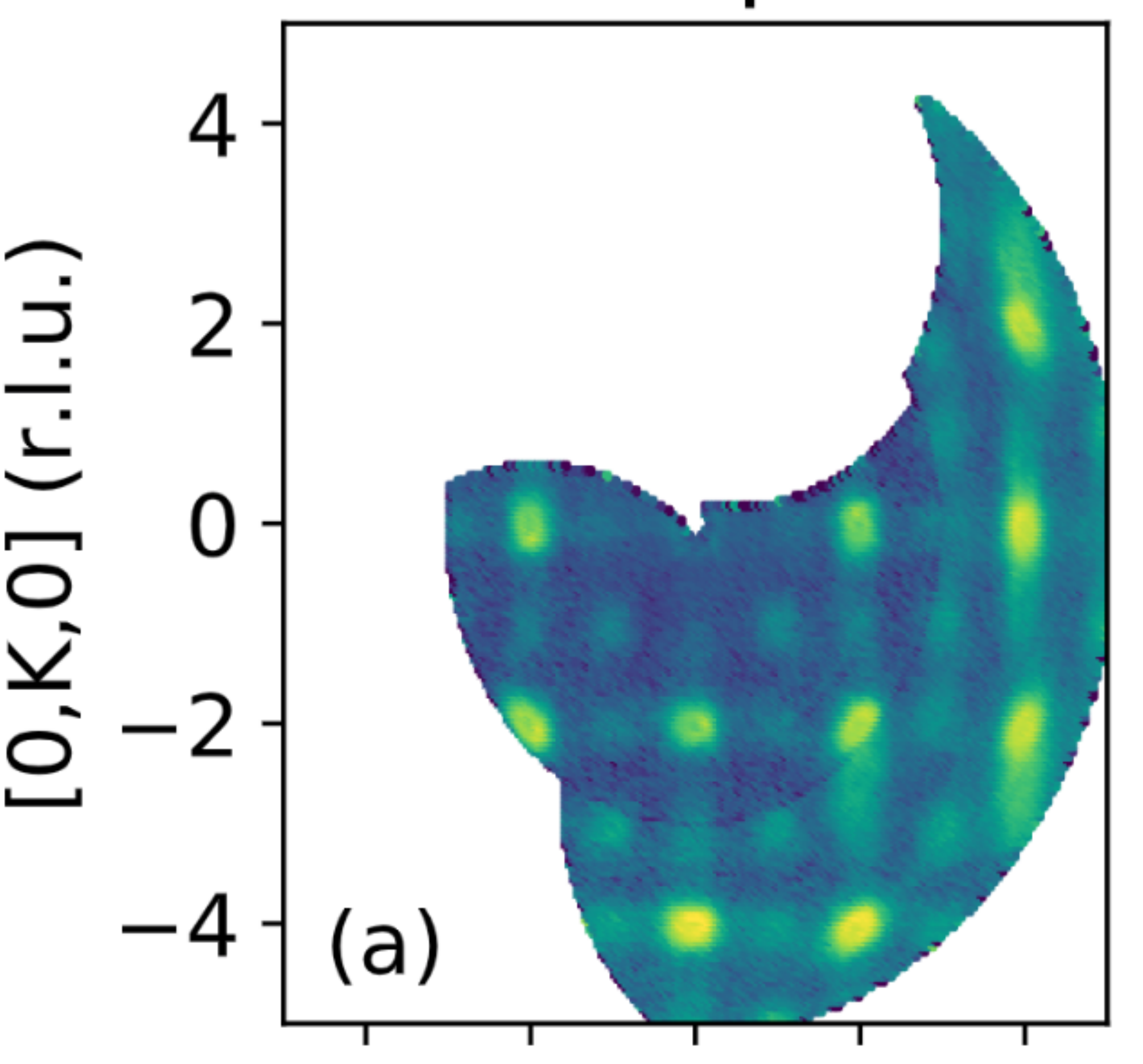}};

\draw [arrow] (exp1) -- (MD);
\draw [arrow] (exp2) -- (MD);
\draw [arrow] (expN) -- (MD);
\draw [arrow] (MD) -| node[anchor=west]{SaveMD}(MDFile);
\draw[->, line width=1mm, red]  (MDFile) -- node[anchor=west]{LoadMD}(MDNew);
\draw [arrow] (MDNew) -- node[anchor=west]{SliceView}(slice);
\end{tikzpicture}
\caption{Schematic overview of the multidimensional event workspace, ``MDWorkspace", workflows used at SNS/HFIR using Mantid's ``SaveMD" to save on disk (once) and the identified ``LoadMD" bottleneck operation (executed many times) to retrieve the entire experimental ensemble containing several orientation outputs.}
\label{fig:schematic_workflow}
\end{figure}
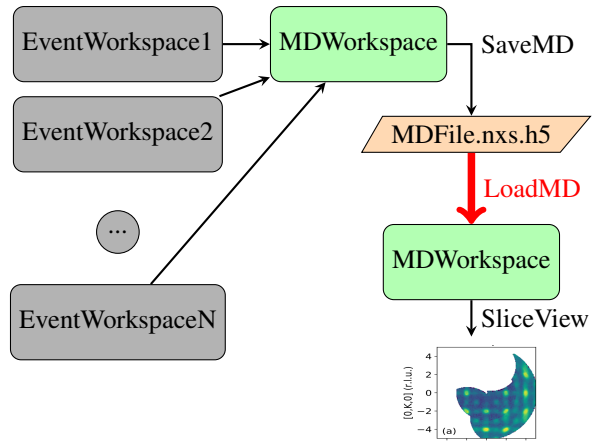

The targeted single-crystal neutron inelastic scattering instruments at SNS and HFIR produce raw event-based data for every experimental setup and conditions. In order to get information on an entire experiment, to construct a meaningful map of the dynamic structure factor $S(\mathbf{Q},E)$, several measurements need to be taken at different sample orientations.  A schematic description is shown in Figure \ref{fig:schematic_workflow} for constructing the resulting data ensemble. We use the Mantid terminology of ``workspaces``~\cite{ARNOLD2014156} for various data structures in memory.   

The raw data for each sample orientation (event workspaces) contain information about neutron detection events (time and position in a detector), together with some metadata about the sample orientation and experiment setup. One transforms this information into a coordinate system relevant to the physics of the materials to be studied, a three dimensional momentum transfer $\mathbf{Q}$ and an energy transfer $E$. This is stored as a multi-dimensional workspace, ``MDWorkspace". Information from multiple sample orientations (or other metadata parameters, such as sample temperature) can be converted to the same coordinates, and stored in the same structure. The resulting ``MDWorkspace" container is then stored in a single HDF5 file using the standard NeXus schema. 

\begin{figure}[!h]
    \centering
    \subfloat[5 experiments]{\includegraphics[width=4.6cm,height=4cm]{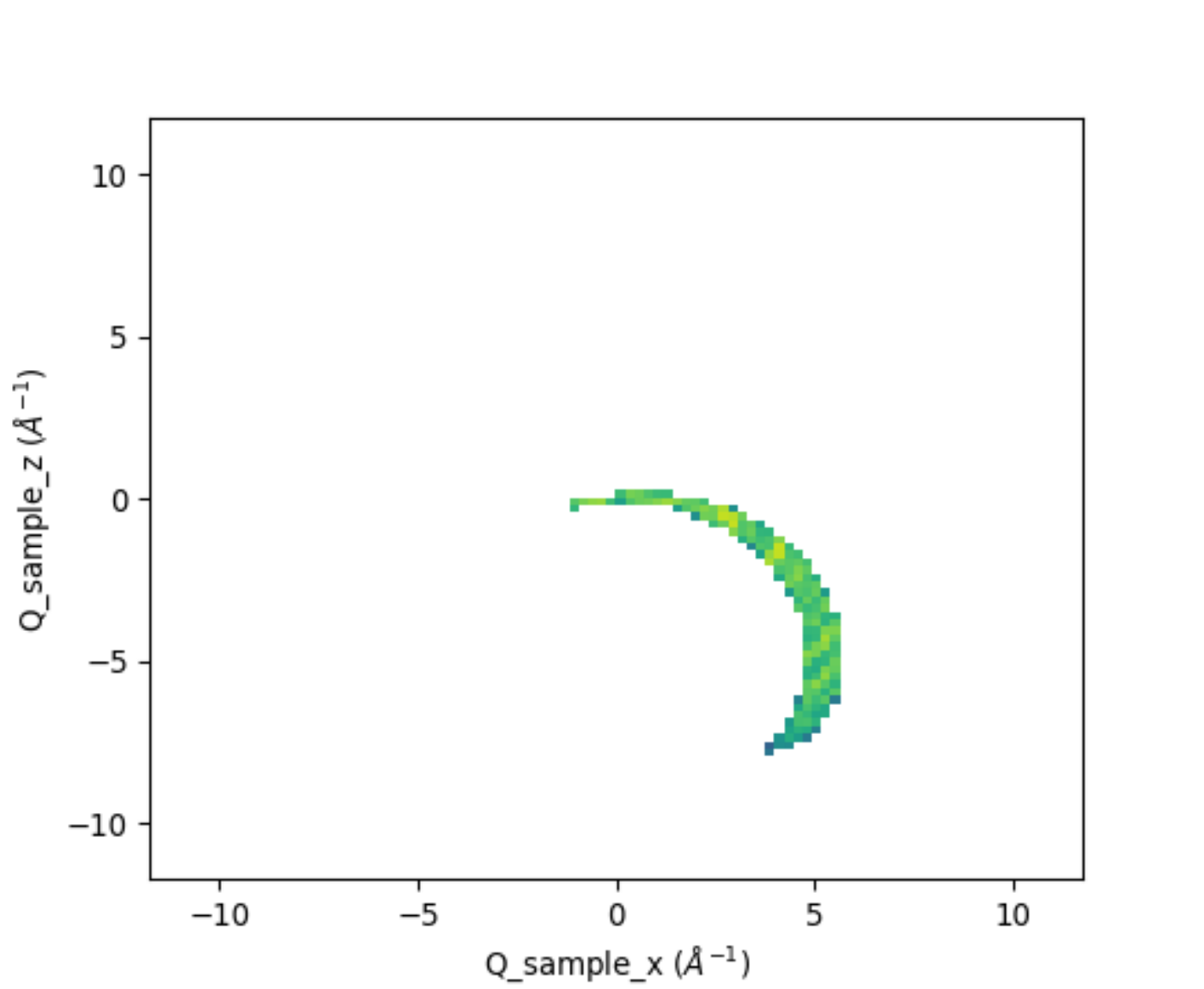}}
    \subfloat[20 experiments]{\includegraphics[width=4.6cm,height=4cm]{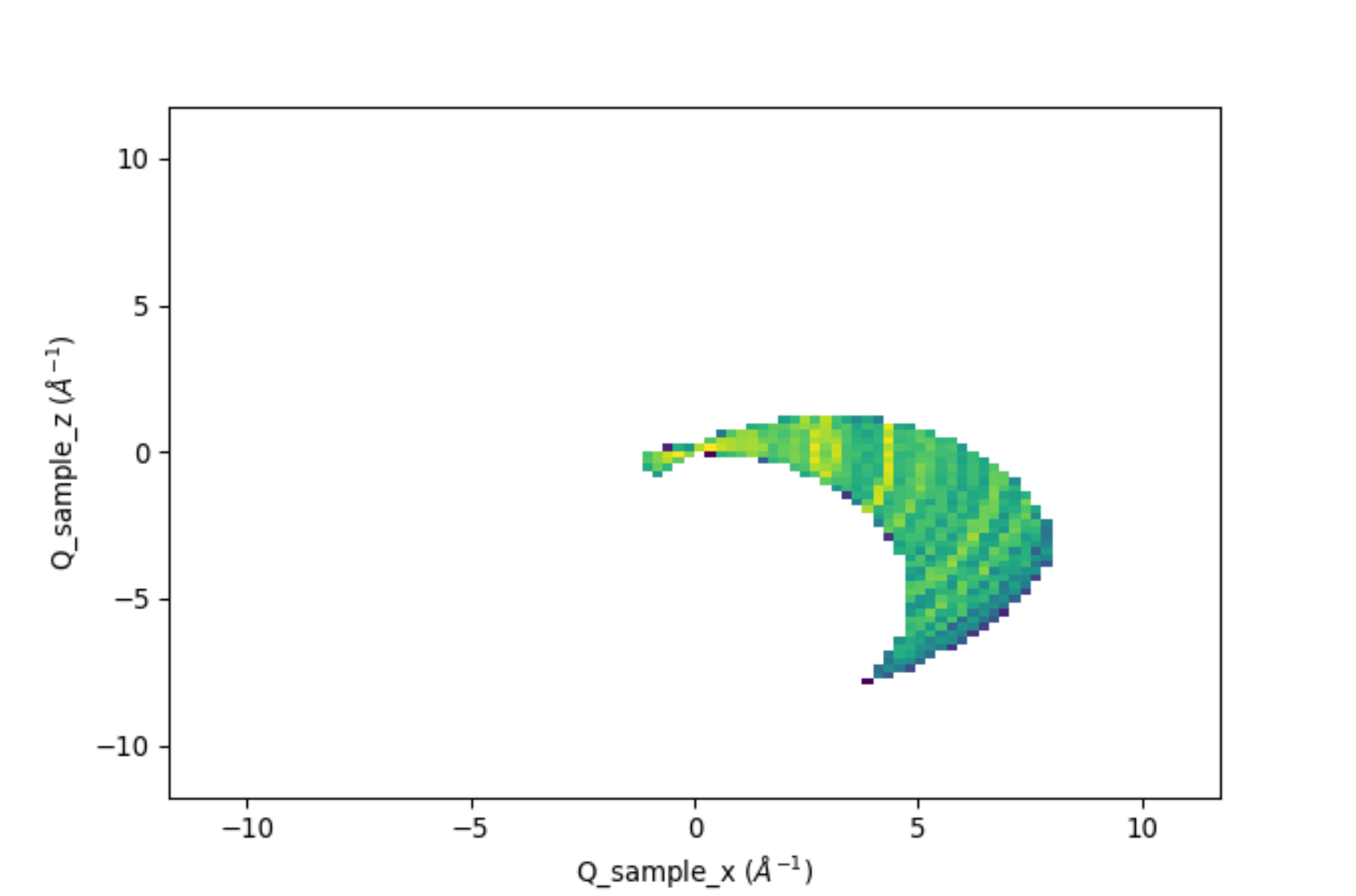}}
    \vspace*{0.001cm} 
    \subfloat[80 experiments]{\includegraphics[width=4.6cm,height=4cm]{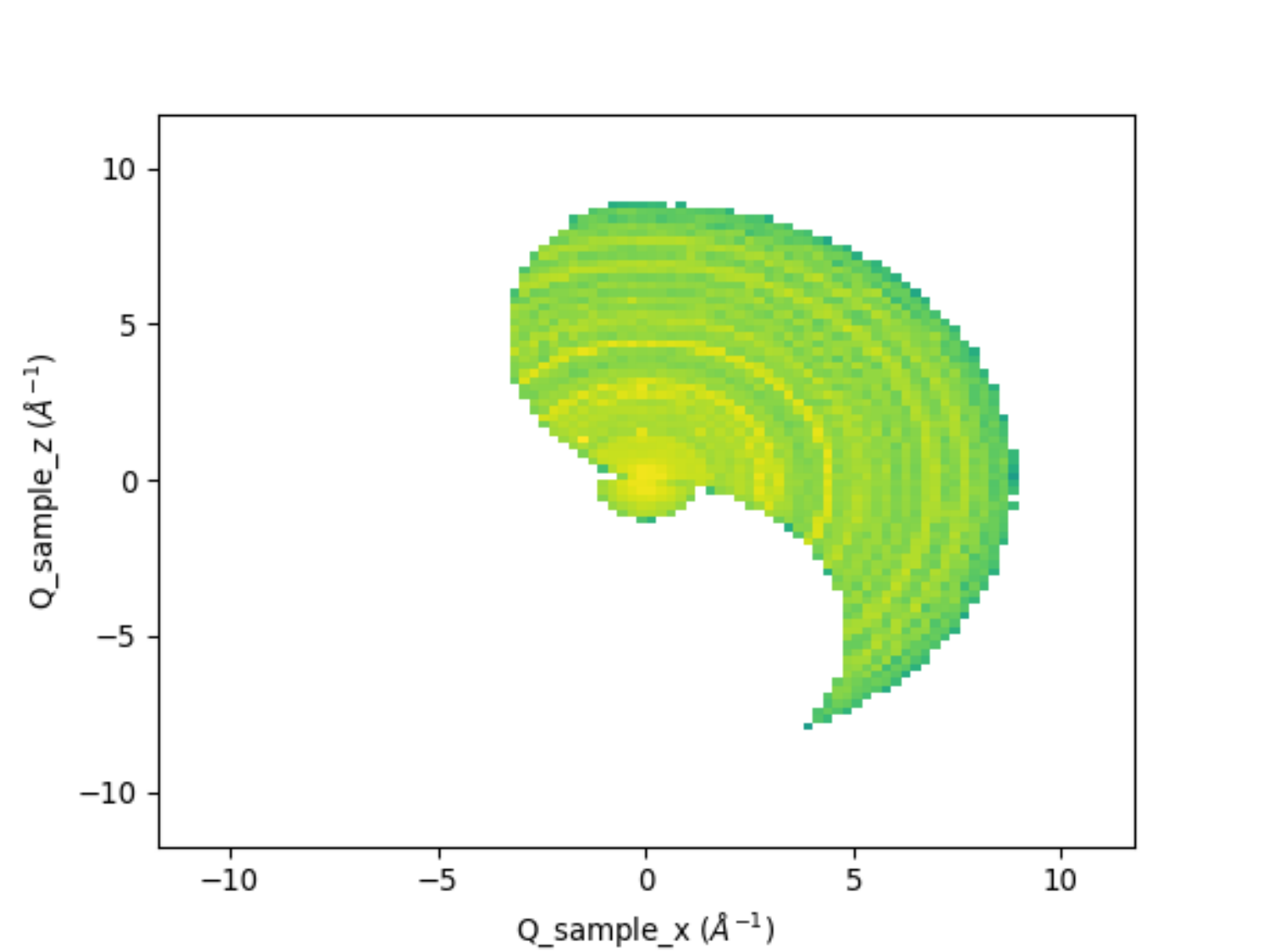}}
    \subfloat[180 experiments]{\includegraphics[width=4.6cm,height=4cm]{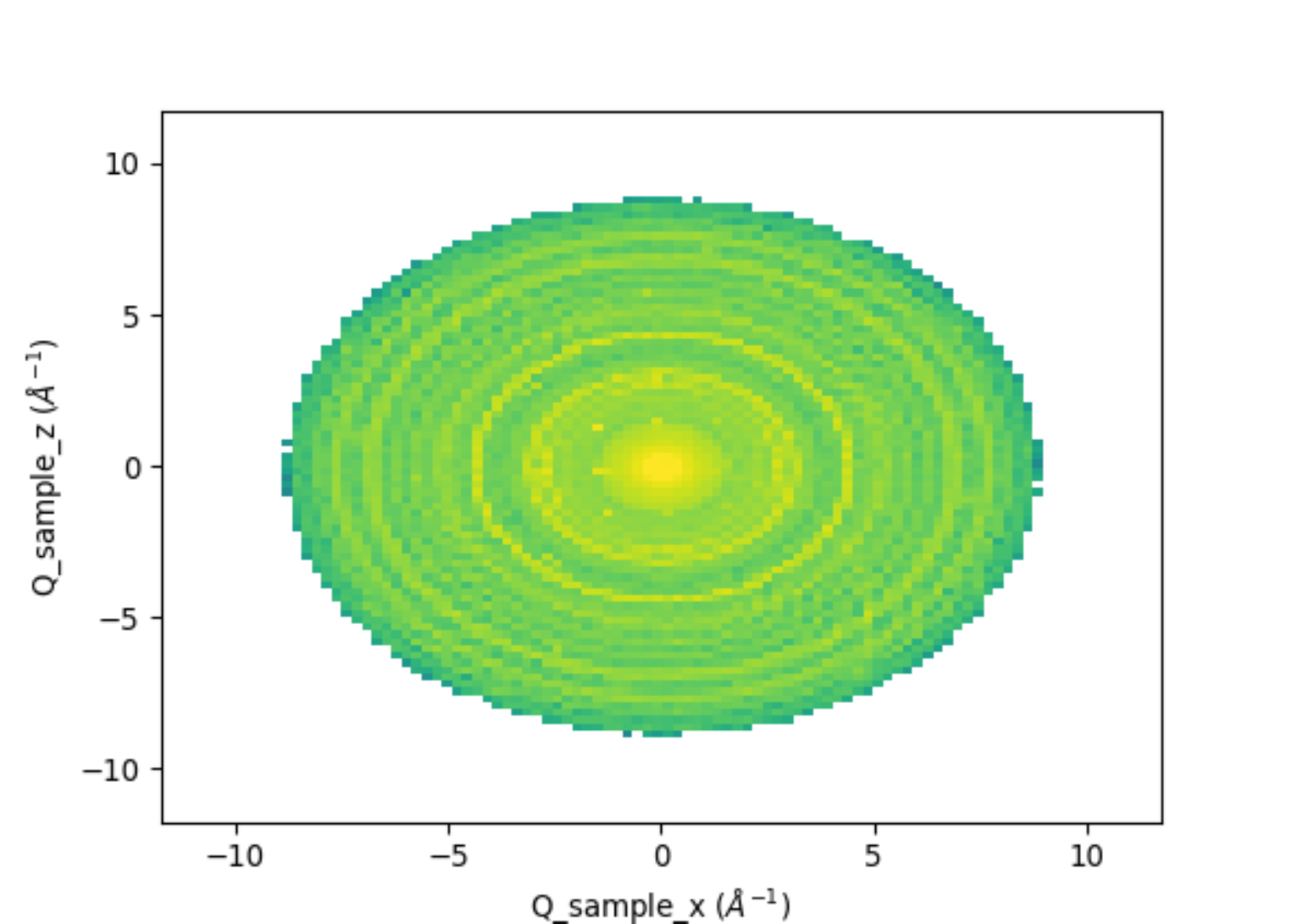}}

\caption{Slice view results of the dynamic structure factor $S(\mathbf{Q}, E)$, as a function of two of the momentum transfer components ($Q_x$ and $Q_z$). The multi-dimensional data is a synthetic one for the ARCS instrument~\cite{doi:10.1063/1.3680104}, covering different orientation ranges of the sample ($2^\circ$ offsets in the sample rotation).}
\label{fig:orientations}
\end{figure}

For visualization and further analysis purposes, researchers typically take multiple one or two dimensional slices out of this workspace, aligned along some arbitrary directions, relevant to the particular sample they study. Pre-processing the raw data into the ``MDWorkspace" in Fig.~\ref{fig:schematic_workflow} is a computationally expensive process, but it is enough to perform it only once. The slicing operation occurs multiple times, with different slicing parameters. It is therefore desirable to store the intermediate data and load it several times using the ``SaveMD'' and ``LoadMD'' algorithms in Mantid. Typical 2D slices for various number of sample orientations are shown in Fig.~\ref{fig:orientations}.

The entries from a typical file ensemble are illustrated in Table~\ref{tab:nexus}. Thus showing the scalable nature of the file size, hence operations, as a function of the number of experiments conducted and stored at the targeted SNS/HFIR instruments. For example, based on the current number of entries listed in Table~\ref{tab:nexus}, a file containing 180 measurements can contain close to 200K entries. This leads to a scalability challenge when processing the metadata and data in ``LoadMD".  

\begin{table}[!h]
\begin{center}
\begin{tabular}{ l l }
Data Type  & Entry Name \\
\hline
group      & /entry \\
attribute  & /entry/NX\_class \\ 
           & ...                \\
group      & /MDEventWorkspace/experiment0 \\
group      & /MDEventWorkspace/box\_structure \\
dataset    & /MDEventWorkspace/coordinate\_system \\
dataset    & /MDEventWorkspace/dimensions \\
group      & /MDEventWorkspace/event\_data \\
           & ... \\
attribute  & /MDEventWorkspace/experiment0/NX\_class \\
group      & /MDEventWorkspace/experiment0/instrument \\
           & ... 20 entries                \\
group      & /MDEventWorkspace/experiment0/logs \\
           & ... 700 to 1000 entries \\
group      & /MDEventWorkspace/experiment0/sample \\
           & ... 17 entries \\
group      & /MDEventWorkspace/process \\
           & ... \\
           & ... \\
           & ... \\
group      & /MDEventWorkspace/experimentN \\
           & ...    
\end{tabular}
\end{center}
\caption{Schematic representation of the hierarchical NeXus schema~\cite{Konnecke:po5029} for ``MDWorkspace" files storing information for several experiments including instrument configuration information (logs, sample, process). Index contents scale up with the total number of experiments (N+1).}
\label{tab:nexus}
\end{table}

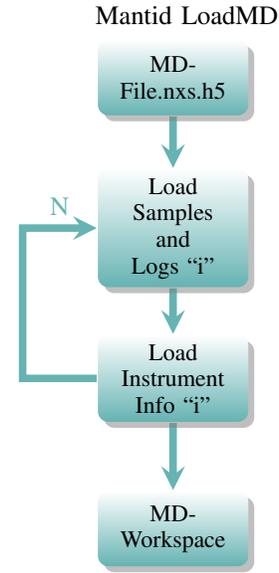
\begin{figure}[!t]
\begin{center}

Mantid LoadMD

\begin{center}

\smartdiagramset{border color=none, uniform color list=teal!60 for 4 items, arrow style=<-, module y sep=2.00, back arrow disabled,}

\smartdiagramadd[flow diagram:vertical]{MDFile.nxs.h5, Load Samples and Logs ``i", Load Instrument Info ``i", MDWorkspace}{}

\begin{tikzpicture}[overlay]
\draw[additional item arrow type,color=teal!60] (module3) -| ++(-2,2) -- node[anchor=south]{N}(module2);
\end{tikzpicture}

\end{center}

\end{center}
\caption{Mantid's ``LoadMD" loop steps for processing entries of a single input ``MDWorkspace" NeXus file containing a ``N" number of experiments.}
\label{fig:LoadMD}
\end{figure}

As expected from being the inverse operation of ``SaveMD", ``LoadMD" essentially retrieves the stored information by looping through the experiments data and metadata from the overall ensemble. This is illustrated in Fig.~\ref{fig:LoadMD} showing the steps for loading samples and log info for each experiment configuration in the dataset ensemble. While, multithreaded parallelization strategies could be an option to speed up the present loop, it is still desired to load the entire experimental data ensemble. It is expected that more complex slicing patterns than those shown in Fig.~\ref{fig:orientations} will not necessarily match the data ordering in the in-memory ``MDWorkspace" data structure. As a result, focusing on the repeated functionality will guide  initial efforts to locate bottlenecks that scale with the number of experiments stored in a single ensemble.   

\section{Proposed Methodology} \label{sec:proposed_methodology}
Similar to our previous effort in speeding up the loading of raw neutron scattering data~\cite{9377836}, we first understand the current bottlenecks in ``LoadMD" due to existing access patterns previously identified in Mantid~\cite{10.1007/978-3-030-63393-6_12}. 

\begin{figure}[!h]
\centering
\subfloat[]{\includegraphics[width=9cm]{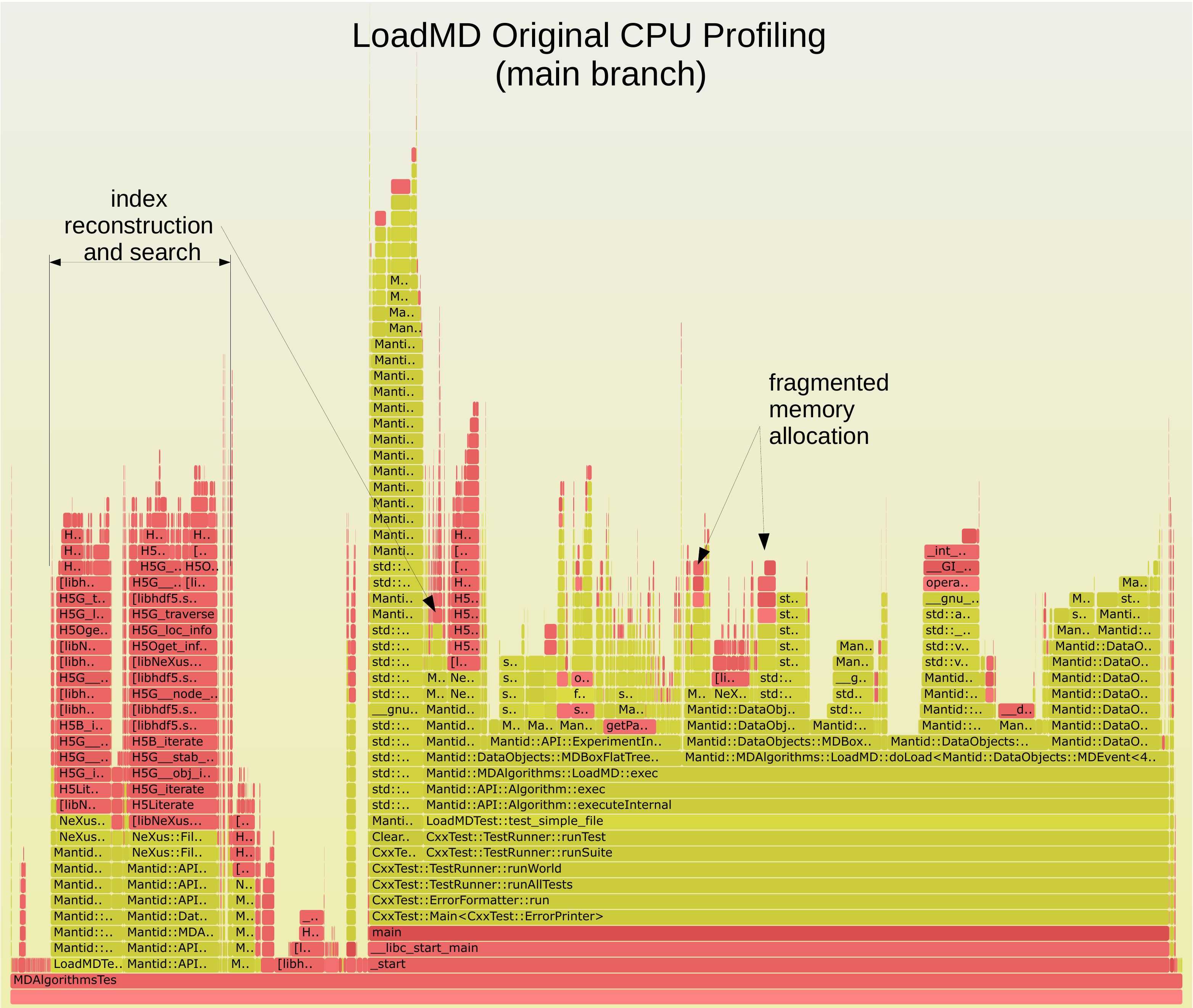}
\label{fig:prof_main}}
\hfil
\hfil
\subfloat[]{\includegraphics[width=9cm]{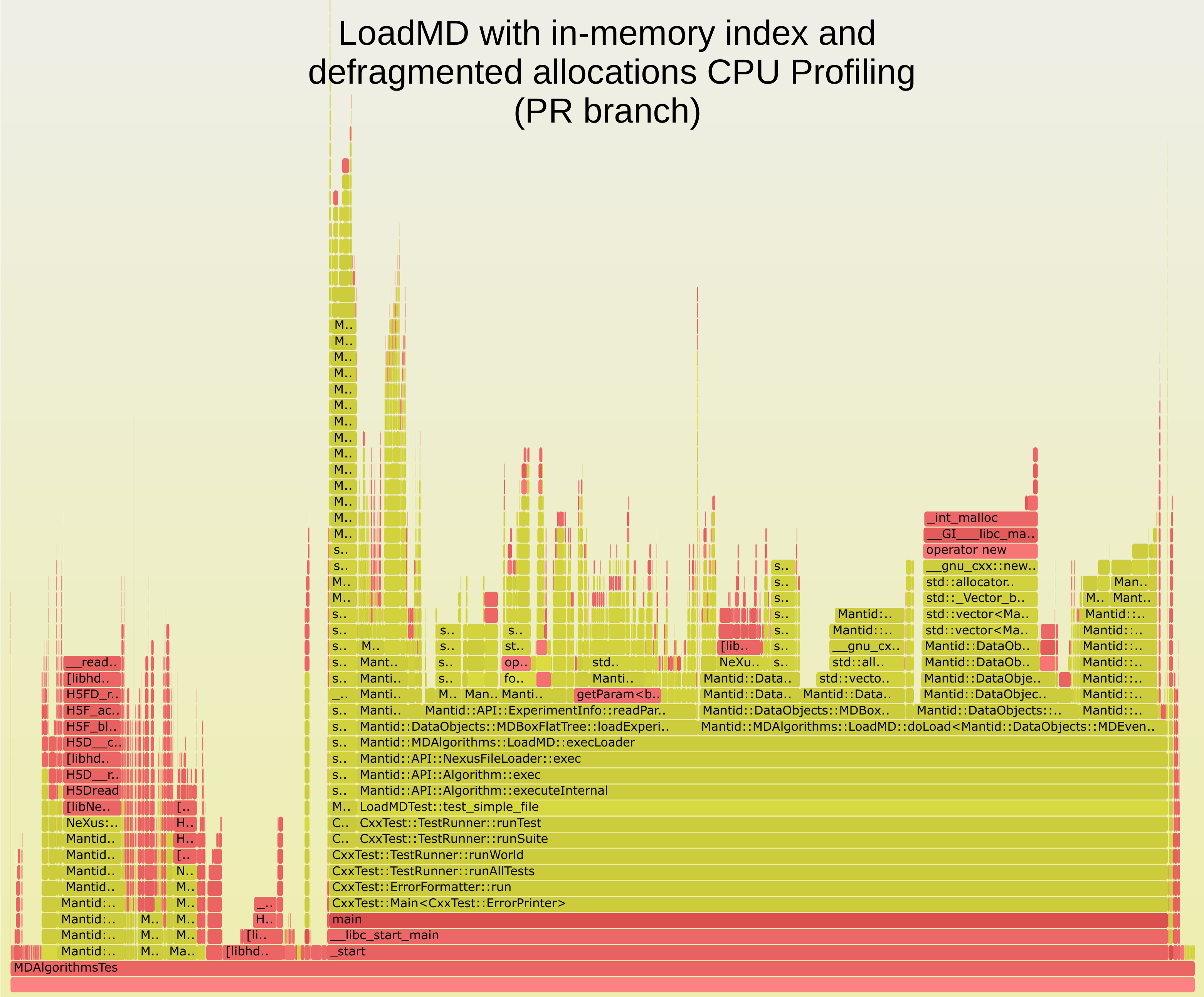}
\label{fig:prof_PR}}
\caption{Mantid's ``LoadMD" CPU profiling flame graph representation for (a) Mantid v6.0 main branch, and (b) proposed PR branch\textsuperscript{\ref{note2}} on Mantid. The reduction of metadata-related CPU operations bottlenecks and memory allocation fragmentation is illustrated.}
\label{fig:prof}
\end{figure}

Figure~\ref{fig:prof} illustrates the result of profiling ``LoadMD" from the existing algorithm in Mantid's main branch, as of October 2021, using flame graphs~\cite{10.1145/2909476} for visualization purposes. The $x$ axis indicates the relative proportion of time for functions inside ``LoadMD" in which most time on the central processing unit (CPU) is consumed, while the $y$ axis indicates the call stack down to the underlying libraries outside Mantid (in red). Figure~\ref{fig:prof_main} shows that functions associated with reconstructing the ``in-memory" metadata index can take a large portion of CPU runtime ($\sim$ 20\%), in addition a few cases of memory reallocation per experiment entry have been identified for potential refactoring that lead to savings in memory operations ($\sim$ 4-5\%). The latter becomes important in the targeted shared resource as requests can add to the operating system (OS) overhead for finding available resources ({\it e.g.} contiguous memory). 

To address the bottleneck due to in-memory index metadata reconstruction operations we reuse the binary-tree formulation described in our previous work~\cite{9377836}. This binary-tree is cached in memory from the beginning of ``LoadMD" and is kept persistent through the loop operations described in Fig.~\ref{fig:LoadMD}. Table~\ref{tab:IndexMemory} shows a schematic representation of the proposed in-memory index for the reduced ensemble stored in a single HDF5 NeXus file with ``SaveMD". Rather than following a hierarchical approach as the data outlined on disk, the entries are sorted by the \texttt{NX\_class} attribute to match processing patterns observed in ``LoadMD" as profiled in Fig.~\ref{fig:prof}. Large data entries are identified as scientific datasets (SDS) and compose the majority of the entries. The final number of SDS entries scale up with the number of experiments stored in a single file.

\begin{table}[!h]
\begin{center}
\begin{tabular}{ l l }
Key: NX\_class  & Value: Sorted binary-tree with absolute-path entry key\\
\hline
NXdata & /MDEventWorkspace/box\_structure \\
       & /MDEventWorkspace/event\_data \\
       & /MDEventWorkspace/experiment0/sample/material \\
 	   & /MDEventWorkspace/experiment1/sample/material \\
       & ... \\
       & /MDEventWorkspace/experimentN/sample/material \\
       & \\
NXentry & /MDEventWorkspace \\
        & \\
NXgroup & /MDEventWorkspace/experiment0 \\ 
        & /MDEventWorkspace/experiment0/logs \\
        & ... \\
        & /MDEventWorkspace/experimentN \\ 
        & /MDEventWorkspace/experimentN/logs \\
        & .. \\
NXinstrument & /MDEventWorkspace/experiment0/instrument \\ 
        & ... \\
        & /MDEventWorkspace/experimentN/instrument \\
        & \\
NXpositioner & /MDEventWorkspace/experiment0/goniometer \\ 
        & ... \\
        & /MDEventWorkspace/experimentN/goniometer \\
        & \\
NXlog   & /MDEventWorkspace/experiment0/logs/gd\_prtn\_chrg \\
        & ... \\
        & /MDEventWorkspace/experimentN/logs/gd\_prtn\_chrg \\
        & \\
SDS &   /MDEventWorkspace/experiment0/instrument/... \\
    &   /MDEventWorkspace/experiment0/logs/... \\
    &   /MDEventWorkspace/experiment0/sample/... \\
    &    ... \\
    &   /MDEventWorkspace/experimentN/instrument/... \\
    &   /MDEventWorkspace/experimentN/logs/... \\
    &   /MDEventWorkspace/experimentN/sample/...\\
    &    ... 
\end{tabular}
\end{center}
\caption{Schematic ``in-memory" binary-tree index from MDFile NeXus file entries implemented using using C++'s \texttt{map<string,set<string>} data structure in our refactoring. Scientific data set (SDS) entries form the largest sub-tree.}
\label{tab:IndexMemory}
\end{table}

A new profiling flame graph is shown in Fig.~\ref{fig:prof_PR} after applying the proposed algorithmic changes to reduce the identified bottlenecks in ``LoadMD".
It can be seen that the index reconstruction functions have been replaced by the proposed cached large binary-tree. Also, refactoring memory allocations and deallocations in fewer operations lead to a less fragmented profile for memory operations. As a result, the redesigned ``LoadMD" algorithm is expected to make fewer calls to the underlying HDF5 library binary tree search functions and the OS ``malloc/free" functions for memory management. The expectation is that these algorithmic changes would impact the overall wall-clock times occupied by users of SNS/HFIR experimental facilities and computational resources\textsuperscript{\ref{note1}}.

\section{Performance Results and Impact} \label{sec:results}

This section presents the performance gains that result from our proposed implementation in Section~\ref{sec:proposed_methodology} on SNS/HFIR production computing systems\textsuperscript{\ref{note1}} that are accessible to users of the neutron science facilities at ORNL~\cite{Campbell_2010}.

After applying the algorithmic changes described in Section~\ref{sec:proposed_methodology}, we compare the performance of ``LoadMD" for the current Mantid ``main" branch and our proposed implementation, accessible through a pull request (PR)\footnote{\label{note2}\url{https://github.com/mantidproject/mantid/pull/32529}}, for a wide range of files with different number of experiments stored with ``SaveMD". These files are described in Table~\ref{tab:mdfiles} showing the typical ranges for the synthetic database resembling products from the Wide Angular-Range Chopper Spectrometer, ARCS instrument~\cite{doi:10.1063/1.3680104} at SNS. It can be seen that both, metadata entries and data sizes, scale up perfectly linearly since the number of stored parameters and data per experiment are fixed when selecting a particular instrument. Nevertheless, these numbers might change as we move to other instruments, which is currently out of the scope of the present work.  

\begin{table}[!h]
\begin{center}
\begin{tabular}{ r r r }
Number of   & Number of  & File Size \\
Experiments & Entries    &  (GB) \\
\hline
10  & 11,000 & 2.1 \\
40  & 43,910 & 8.1 \\ 
80  & 88,887 & 17.0 \\
180 & 198,587 & 37.0 
\end{tabular}
\end{center}
\caption{Description of the synthetic file ensembles based on the ARCS instrument~\cite{doi:10.1063/1.3680104} products used for the performance and scalability tests in this study.}
\label{tab:mdfiles}
\end{table}

\begin{figure}[!h]
    \centering
    \subfloat[40 experiments]{\includegraphics[width=8cm,height=5cm]{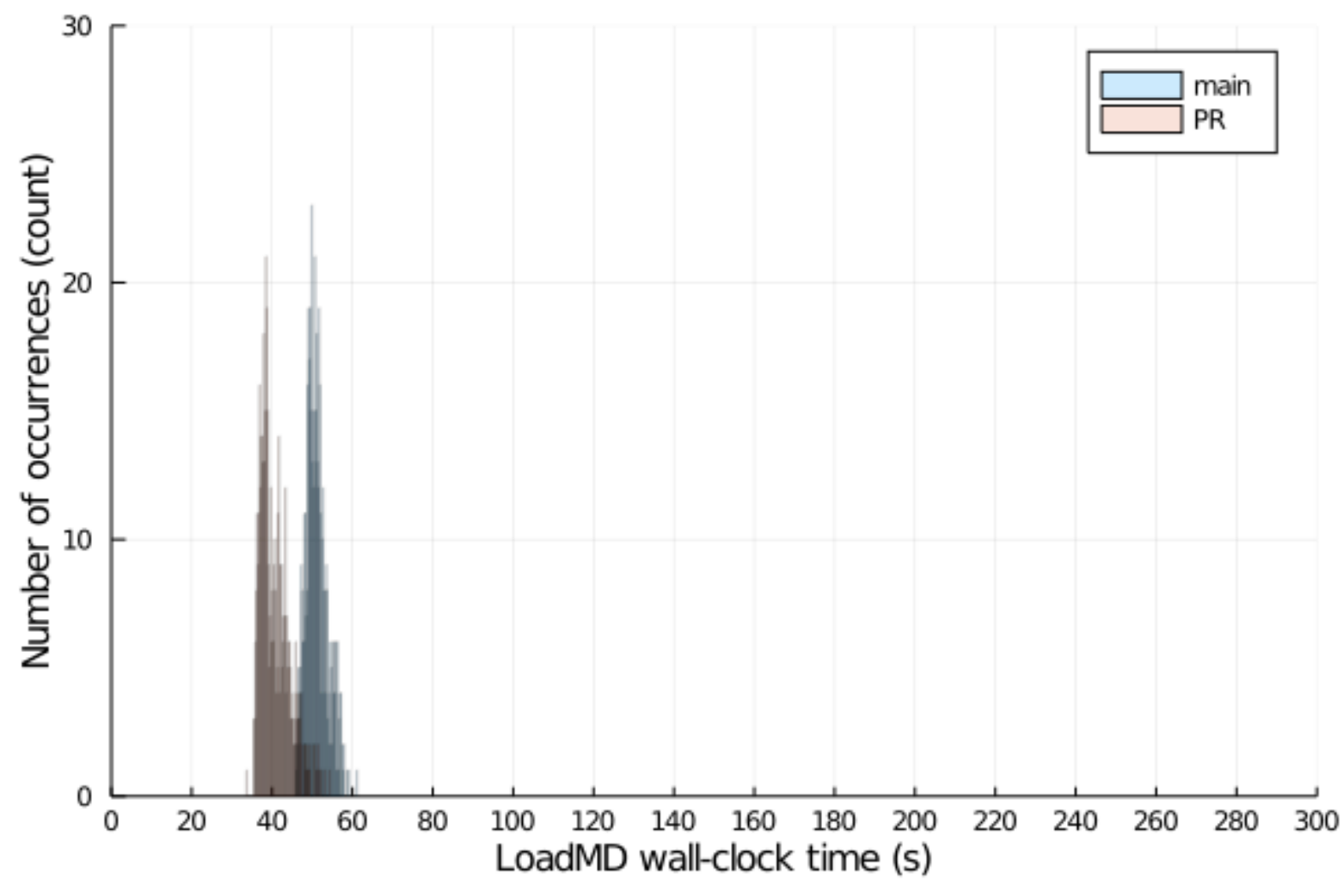}}
    \vspace*{0.001cm} 
    \subfloat[80 experiments]{\includegraphics[width=8cm,height=5cm]{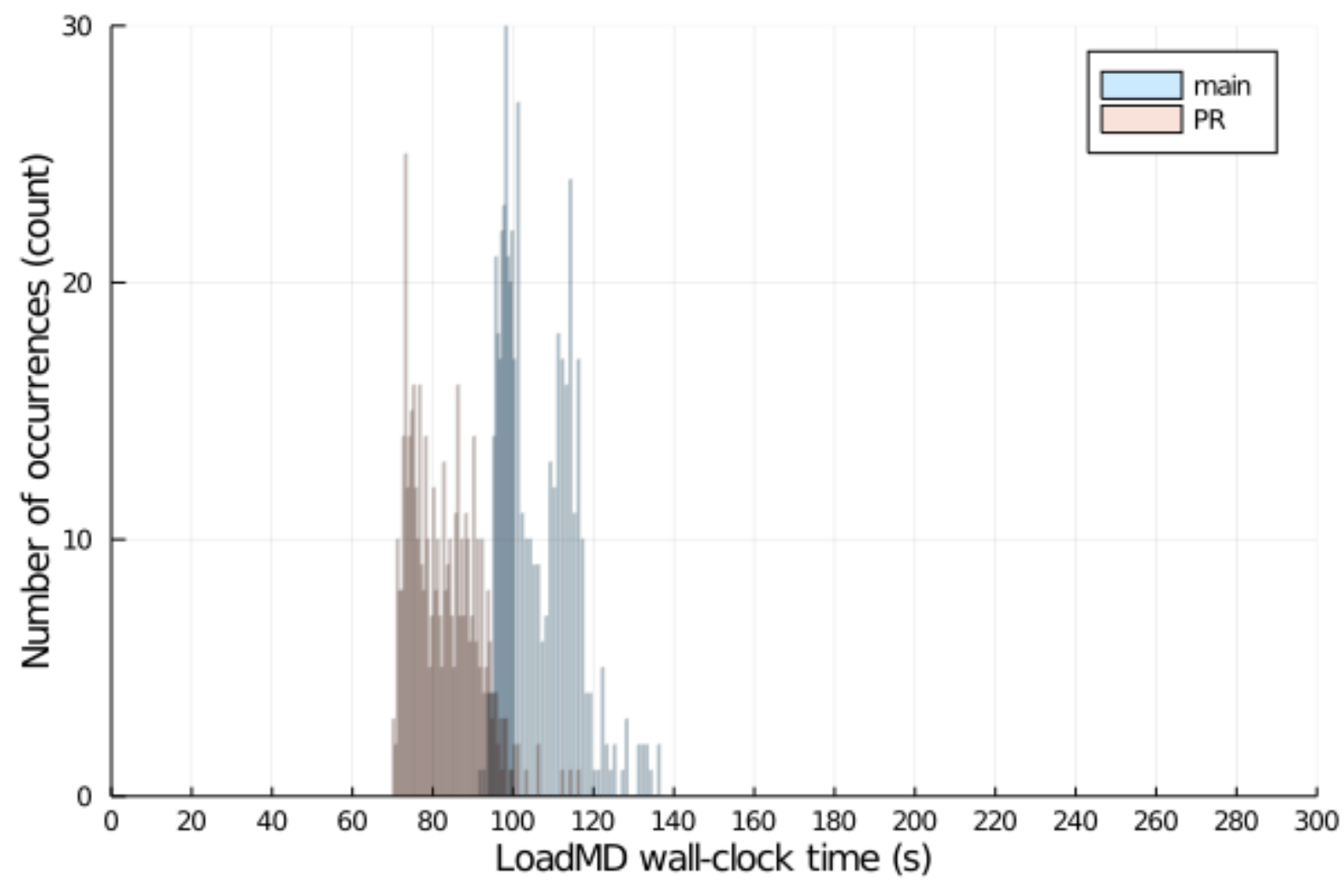}}
    \vspace*{0.001cm} 
    \subfloat[180 experiments]{\includegraphics[width=8cm,height=5cm]{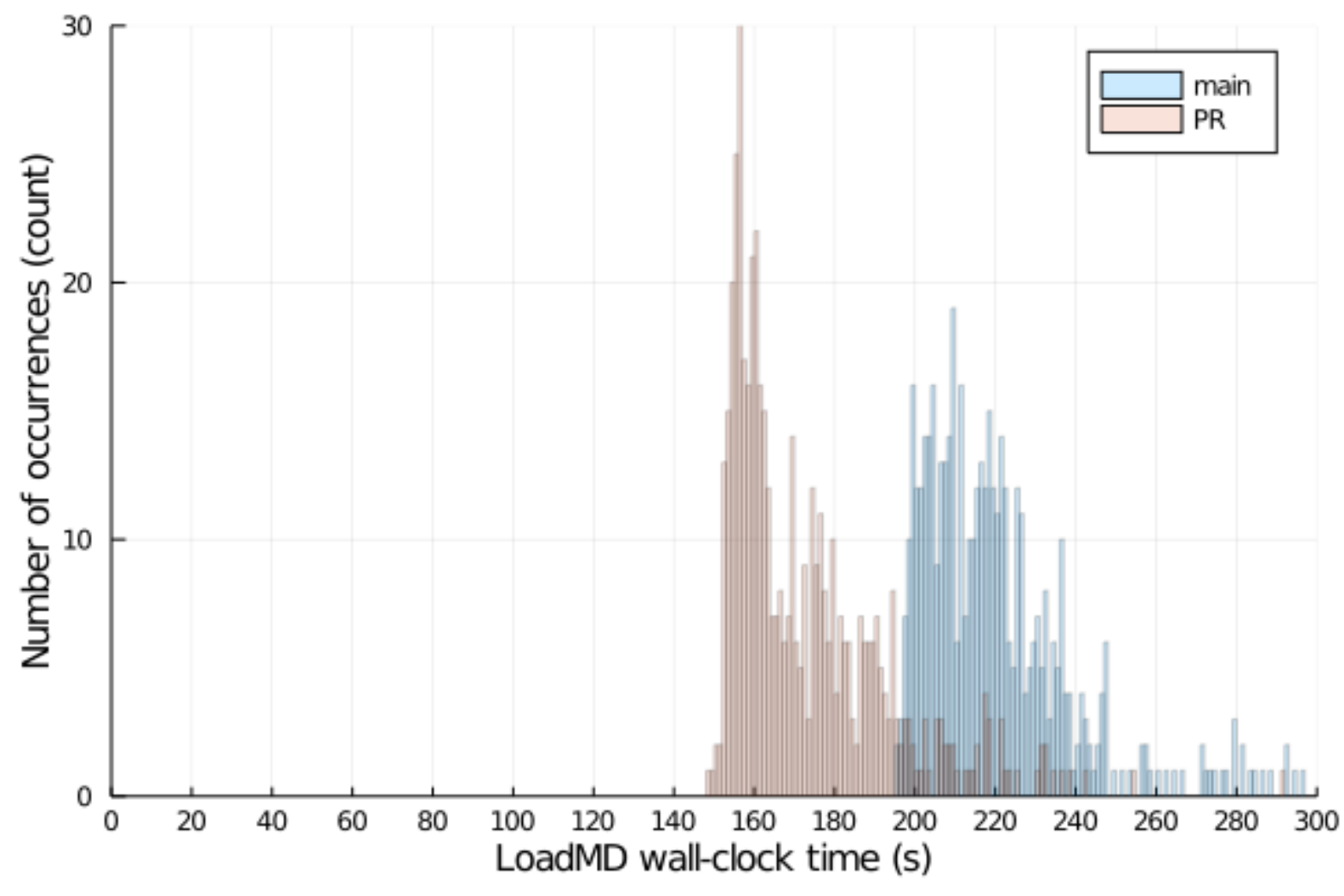}}
\caption{Wall-clock times histograms for current Mantid ``main" branch and the speed ups proposed changes from our pull request (PR), showing a 19\% to 23\% improvements on average for all cases on SNS/HFIR computing systems\textsuperscript{\ref{note1}}.}
\label{fig:results_histogram}
\end{figure}

To obtain a meaningful benchmark statistic, we ran 500 computational experiments for each case in Table~\ref{tab:mdfiles} on a single node on SNS/HFIR computational systems\textsuperscript{\ref{note1}}.
Each node is powered by a 48-core Intel Xeon E5-2670 CPU and 512\,GB of random-access memory (RAM). It is important to mention that system variability must be considered in our results due to the shared nature of the resources, thus rather than a deterministic analysis we obtain a distribution from our benchmarks tests capturing overall ``LoadMD" wall-clock times. The obtained wall-clock histograms are presented in Fig.~\ref{fig:results_histogram} for the largest cases in Table~\ref{tab:mdfiles}, (40, 80 and 180 experiments). Overall, it can be seen that even when system variability is considered, the modifications introduced in this work result in consistent speed ups for any number of experiments. As expected, Fig~\ref{fig:results_histogram} also shows that system variability has a noticeable influence on ``LoadMD" wall-clock times as the stored number of experiments, thus operations, increase per target file. The system variability is more likely to be influenced by the state of the system when performing this calculations: CPU load, memory availability, file disk requests. Hence, we observe the effect that reducing the number of requests to lower-level memory and I/O operations also reduces the variability introduced by other users requesting system resources.

\begin{figure}[!h]
\centering
\includegraphics[width=10cm,height=6.5cm]{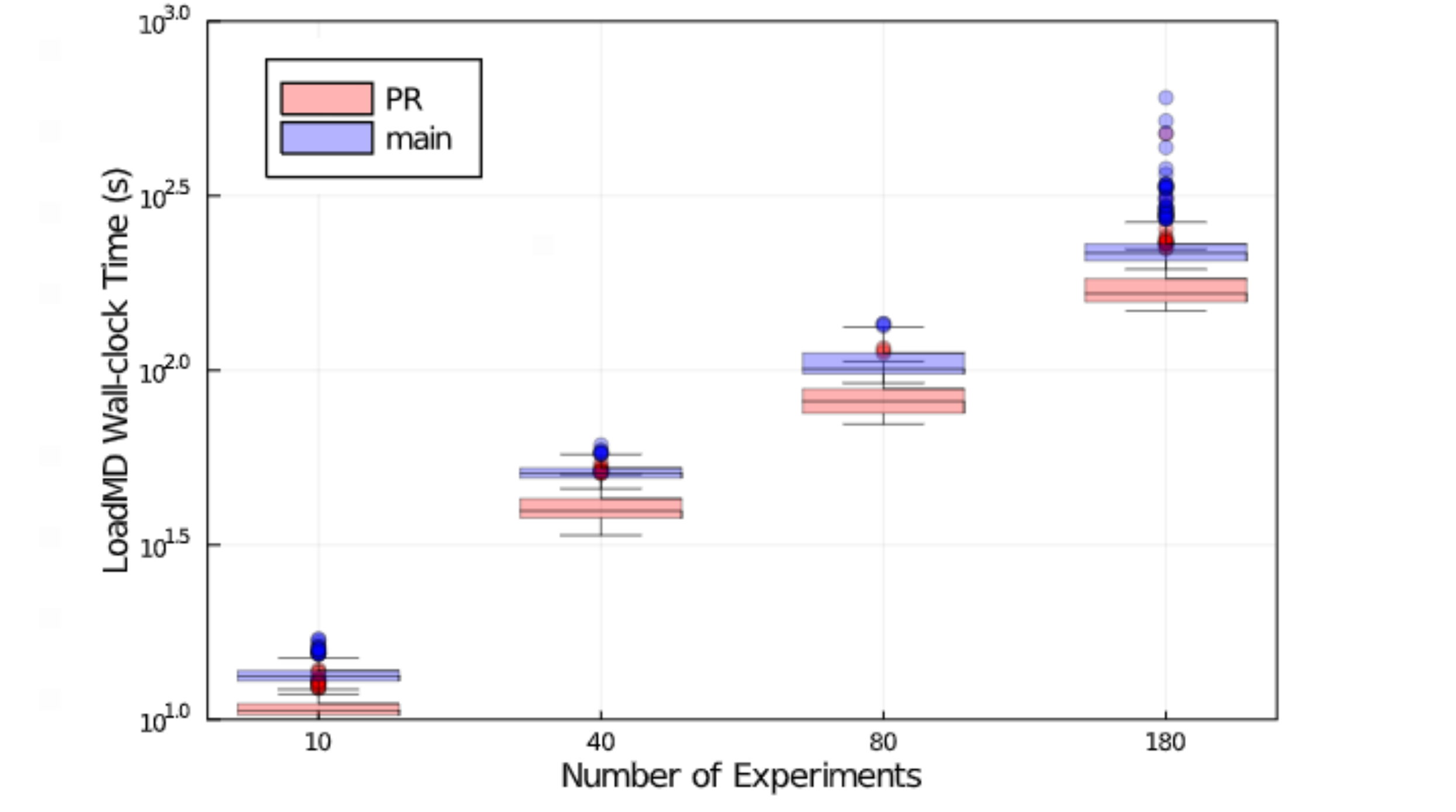}
\caption{Boxplots showing a consistent reduction in the overall wall-clock times for Mantid's ``LoadMD" operation on a single node on SNS/HFIR computational resources\textsuperscript{\ref{note1}} for a wide range of ``MDWorkspace" files with a different number of experiments from SNS/HFIR instruments.}
\label{fig:results_boxplot}
\end{figure}

\begin{table}[!h]
\centering
  \begin{tabular}{|r|r|r|r|r|r|}
    \hline
     Number of &
      \multicolumn{2}{c|}{WC median\,(s)}  &
      \multicolumn{2}{c|}{WC stddev\,(s)}  &
                         {WC median } \\
    Experiments  & main & PR & main & PR & speed up  \\
    \hline
    10 & 13.3 & 10.6 & 0.7 & 0.6 & 20.3\% \\
    \hline
    40 & 50.7 & 39.6 & 2.6 & 3.6 & 21.9\%  \\
    \hline
    80 & 101.0 & 81.5 & 8.7 & 8.1  & 19.2\% \\
    \hline
    180 & 216.0 & 166.0 & 42.1 & 24.8 & 23.1\% \\
    \hline
    360 & 357.0 & 285.0 & 59.4 & 32.9 & 20.3\% \\
    \hline
  \end{tabular}
\caption{Statistics comparing ``LoadMD" wall-clock times from 500 experimental runs on a single node of SNS/HFIR computational resources\textsuperscript{\ref{note1}} using Mantid's main branch and the current PR effort\textsuperscript{\ref{note2}} for different numbers of experiments on a single file.}
\label{tab:summary}
\end{table}

For completeness, we present the statistical box plots showing the consistent speed ups in terms of median values of the obtained wall-clock times on SNS/HFIR systems in Fig.~\ref{fig:results_boxplot}. This includes improvements for even the smaller case for files containing 10 experiments. To further illustrate the overall speed ups, we present a summary of the obtained wall-clocks time median, standard deviation, and speed ups from applying ``LoadMD" on files up to 360 experiments in Table~\ref{tab:summary}. It is interesting to note that while we confirm the speed ups in terms of the median, the standard deviation indicating the dispersion of the measurements might not correlate with the saving in wall-clock times. We infer that this might be due to system interference and more runs might be necessary for capturing different states of the system. Nevertheless, these are measurements worth considering in order to be confident that the present algorithmic improvements will have a very high probability to provide a consistent 19\% to 23\% speed up when compared to the current Mantid ``LoadMD" implementation.

\section{Conclusions} \label{sec:conclusions}
The present work reuses efficient strategies for addressing bottlenecks when processing reduced data ensembles stored using a single HDF5 file and the NeXus schema for a range of experimental setups. The latter is a typical use-case on time-of-flight instruments at ORNL neutron science facilities: SNS and HFIR. Bottlenecks due to index reconstruction and memory allocation and deallocation operations have been identified and addressed on Mantid's ``LoadMD" function that loops through different experimental configurations. ``LoadMD" serves as a backend for loading reduced data ensembles in several reduction workflows that provide facility users with a complete picture of the dynamic structure factor, $S(Q, E)$, that is typical in neutron scattering experiments. Two algorithmic changes on ``LoadMD" are proposed: i) reusing a cached in-memory index for efficient entry search (15-20\% savings), and ii) reducing the number of memory allocations and deallocations in ``LoadMD" (4-5\% savings). The proposed algorithmic changes were tested across files stored at SNS/HFIR covering a typical range for the data ensemble size (10 to 360 experiments).
Tests ran on a single node powered with a 48-core Intel Xeon E5-2670 CPU and 512\,GB of RAM on the SNS/HFIR hosted computational resources available to facility users\textsuperscript{\ref{note1}}.
Results show consistent speed ups in the range of 19\% to 23\% for the median of the measured wall-clock times distributions expected from the shared nature of the resources.
The proposed improvements have been recently integrated into the Mantid open-source framework, thus expecting further impact from our work in other neutron science facilities using ``LoadMD" for retrieving data ensembles from multiple experiment configurations.

\section*{Acknowledgment}
We would like to thank B. Ueland, B. Li, R. McQueeney, and T. Han for providing the data for testing and benchmarking purposes.
We would like to acknowledge Dr. Chen Zhang for providing feedback to improve the quality of the manuscript.
This manuscript has been authored by UT-Battelle, LLC under Contract No. DEAC05-00OR22725 with the U.S. Department of Energy. The United States Government retains and the publisher, by accepting the article for publication, acknowledges that the United States Government retains a nonexclusive, paid-up, irrevocable, world-wide license to publish or reproduce the published form of this manuscript, or allow others to do so, for United States Government purposes. The Department of Energy will provide public access to these results of federally sponsored research in accordance with the DOE Public Access Plan (\url{http://energy.gov/downloads/doe-public-access-plan}).
Work at Oak Ridge National Laboratory was sponsored by the Division of Scientific User Facilities, Office of Basic Energy Sciences, US Department of Energy, under Contract no. DE-AC05-00OR22725 with UT-Battelle, LLC.

\ifCLASSOPTIONcaptionsoff
  \newpage
\fi

\bibliographystyle{IEEEtran}
\bibliography{IEEEabrv,paper.bib}

\end{document}